\documentclass[english,aps,tightenlines,showpacs, showkeys, notitlepage]{revtex4-2}
\usepackage[T1]{fontenc}
\usepackage[latin9]{inputenc}
\usepackage{color}
\usepackage{babel}
\usepackage{amsbsy}
\usepackage{amstext}
\usepackage{setspace}
\usepackage{esint}
\usepackage[unicode=true,pdfusetitle,
 bookmarks=true,bookmarksnumbered=false,bookmarksopen=false,
 breaklinks=false,pdfborder={0 0 0},pdfborderstyle={},backref=false,colorlinks=true]
 {hyperref}
\begin{document}
\title{Modified Friedmann Equations from Maxwell-Weyl Gauge Theory}
\author{Salih Kibaro\u{g}lu}
\email{salihkibaroglu@maltepe.edu.tr}

\date{\today}
\begin{abstract}
This study investigates the possibility of a homogeneous and isotropic
cosmological solution within the context of the Maxwell-Weyl gauge
theory of gravity. To achieve this, we utilize the Einstein-Yang-Mills
theory as an analogy and represent the Maxwell gauge field in terms
of two time-dependent scalar fields. We derive the modified Friedmann
equations, integrating the contributions from the Maxwell gauge fields
and an effective cosmological constant that depends on the Dirac scalar
field. Our analysis reveals how these modifications influence various
cosmological scenarios, including power-law evolution, de Sitter-like
expansion, inflationary phases, non-singular bounce cosmologies, and
cyclic cosmologies. 
\end{abstract}
\affiliation{Maltepe University, Faculty of Engineering and Natural Sciences, 34857,
Istanbul, Turkey}
\affiliation{Institute of Space Sciences (IEEC-CSIC) C. Can Magrans s/n, 08193
Barcelona, Spain}
\pacs{02.20.Sv, 11.15.-q, 98.80.-k, 95.36.+x}
\keywords{Cosmology, Dark energy, Gauge Field Theory, Lie algebra, }
\maketitle

\section{Introduction}

The general theory of relativity, proposed by Albert Einstein, is
a fundamental theory of gravity that forms the basis for our current
understanding of cosmology. One of the most successful theories in
cosmology is the $\Lambda$CDM model which describes the universe
as being composed of dark matter, dark energy, and baryonic matter.
It assumes that the universe is homogeneous and isotropic on large
scales and that it began with a hot and dense Big Bang. However, there
are several reasons we need to generalize current cosmological models.
Addressing some of the most fundamental questions in physics, such
as the nature of dark matter and dark energy, the origin and evolution
of the universe, and the potential existence of other universes or
dimensions beyond our own. By exploring these questions and developing
more sophisticated models, we can gain deeper insights into the nature
of reality and our place in the cosmos.

In the 1950s and 1960s, physicists such as Yang-Mills, Utiyama, and
Kibble began exploring the application of gauge theory to the study
of fundamental forces, including gravity \citep{Yang:1954Conservation,Utiyama:1956Invariant,Kibble:1961LorentzInvariance}.
They developed the concept of gauge invariance, which played a key
role in the development of the Standard Model of particle physics
and lead to develop the concept of the gauge theory of gravity. The
gauge theory of gravity provides a powerful framework to reach a more
comprehensive gravitational model by extending the gravitation theory
to include additional fields and symmetries. In this framework, gravity
is described as a curvature of space-time, and the force of gravity
is seen as a manifestation of the curvature. This has important implications
for cosmology, as it implies that the large-scale structure of the
universe is intimately connected to the curvature of space-time. For
example, the gauge theory of gravity, which is based on the Weyl group,
constitutes a potent framework for understanding the properties and
behavior of the universe on its largest scales, and scale invariance
plays a key role in this framework \citep{Bregman:1973Weyl,Charap:1973GaugeWeyl,Dirac:1973LongRange,Kasuya:1975Weyl,Rosen:1982Weyl}.
The scale invariance is a well-known symmetry \citep{Dirac:1936Wave,Mack:1969Finite}
that has been studied in many cosmological contexts especially in
early and late-time universe models (for recent development, see \citep{Shaposhnikov:2009Scale,Israelit:2011Weyl,Bars:2013LocalConf,Aguila:2014PresentAccelerated,Karananas:2016Scale,Ferreira:2018Inflation,Casas:2019Scale,Ghilencea:2021GaugingScale,Tang:2020WeylSymmetry}).
These studies reveal that in order to gain additional insights into
physical systems, one can use new symmetries or expand the existing
symmetries within the system.

In the pursuit of constructing cosmological models rooted in the gauge
theory of gravity, one of the main challenges is the suitable implementation
of gauge fields. The literature has previously examined gauge field
configurations in the Friedmann-Lemaître-Robertson-Walker (FLRW) cosmology
\citep{cervero1978classical,henneaux1982remarks,galt1991yang,moniz1993dynamics,bamba2008inflationary,gal2008non,maleknejad2013gauge,guarnizo2020dynamical}.
One area of focus is the non-Abelian Yang-Mills (YM) theories, where
the coupling of a non-Abelian YM field to the scalar curvature is
analyzed to determine the cosmological implications of nonminimal
gravitational coupling \citep{maleknejad2011non,sheikh2012gauge,maleknejad2013gauge,bielefeld2015cosmological}
(for a comprehensive review, see \citep{maleknejad2013gauge_report}).
In addition to its use in inflationary cosmology, YM theory is applied
in models for dark energy \citep{Zhao:2005TheStateEquation,gal2008non,bamba2008inflationary,gal2012yang,elizalde2013cosmological,setare2013warm,rinaldi2015dark,mehrabi2017gaugessence}
and dark matter \citep{zhang2010dark,buen2015non,gross2015non}.

An intriguing extension of the theory of gravity can be derived from
the gauge theory of the Maxwell algebra. The Maxwell algebra can be
interpreted as a modified version of the Poincaré algebra, incorporating
six additional tensorial abelian generators that render the four-momenta
non-commutative \citep{bacry1970group,schrader1972maxwell,soroka2005tensor},
\begin{equation}
\left[P_{a},P_{b}\right]=iZ_{ab}.
\end{equation}
The Maxwell algebra provides a useful background for constructing
a gauge theory of gravity, resulting in a generalized theory that
encompasses the cosmological constant and an additional energy-momentum
tensor term \citep{azcarraga2011generalized,durka2011gauged,soroka2012gauge,azcarraga2014minimalsuper,cebeciouglu2014gauge,cebeciouglu2015maxwell,concha2015generalized,kibarouglu2019maxwellSpecial,kibarouglu2019super,kibarouglu2020generalizedConformal,kibarouglu2021gaugeAdS,cebeciouglu2021maxwellMetricAffine,Cebecioglu:2023MaxwellFR,Kibaroglu:2023GLSL}.
This gravitational model is referred to as the Maxwell gauge theory
of gravity or Maxwell gravity, and while the energy-momentum term
remains less explored, on the other hand, it is known that such an
additional term may be related to dark energy \citep{frieman2008dark,padmanabhan2009dark}.
A minimal cosmological model with Maxwell symmetry is introduced in
the literature \citep{durka2011local}, and it is also suggested that
the gauge fields of this symmetry group may provide a geometric background
for describing vector inflatons in cosmological models \citep{Azcarraga2013maxwellApplication}.
Furthermore, there is another study that proposes a cosmological model
for the Maxwell gravity which is constructed by gauging the semi-simple
extended Poincaré algebra \citep{Kibaroglu:2022CosmoMaxwell}.

In this work, we extend our previous study \citep{Kibaroglu:2022CosmoMaxwell}
to include the scale invariance which arises naturally in the Maxwell-Weyl
gauge theory of gravity \citep{cebeciouglu2014gauge} to find more
general cosmological solutions. Our concept is very similar to the
gravitation model based Lyra's geometry \citep{Lyra:1951Modifikation}
which can be seen as a modification of Weyl's geometry. Lyra's gravity
presents a scalar-tensor gravity as a geometrical way in the framework
of Lyra's geometry and the cosmological models based on this geometry
offer a valuable framework for studying the concepts of dark energy
and the cosmological constant \citep{Singh:1993Lyra,Maurya:2019Brans-Dicke-Lyra}.
Just like this model, our objective is to achieve this goal by utilizing
a method that is similar in nature to the one used for the realization
of gauge fields in terms of time-dependent scalar fields. This method
relies on the application of the Einstein-Yang-Mills theory and its
cosmological implications.

The present manuscript is structured as follows. In Section \ref{sec:Maxwell-Weyl-gauge-theory},
we present a concise overview of the Maxwell-Weyl gauge theory of
gravity and the geometric interpretation of the Weyl transformation.
In Section \ref{sec:Cosmological-setup}, we investigate the potential
solutions for the gravitational field equations arising from the framework
of Maxwell-Weyl gravity and give the modified version of the Friedmann
equations. In Section \ref{sec:Applications-for-various}, we discuss
several significant cosmological scenarios under the theoretical framework
of Maxwell-Weyl gauge theory of gravity. Finally, the last section
provides conclusive remarks and discussions.

\section{Maxwell-Weyl gauge theory of gravity\label{sec:Maxwell-Weyl-gauge-theory}}

This section provides a concise overview of the findings from our
prior investigation \citep{cebeciouglu2014gauge}. The mathematical
framework of the Maxwell-Weyl algebra, denoted as $\mathcal{MW}$,
may be regarded as a Weyl extension of the Maxwell algebra that encompasses
a dilatation generator \citep{bonanos2009infinite}.

The non-zero commutation relations of $\mathcal{MW}$ algebra are
given by,

\begin{eqnarray}
\left[M_{ab},M_{cd}\right] & = & i\left(\eta_{ad}M_{bc}+\eta_{bc}M_{ad}-\eta_{ac}M_{bd}-\eta_{bd}M_{ac}\right),\nonumber \\
\left[P_{a},P_{b}\right] & = & i\lambda Z_{ab},\nonumber \\
\left[M_{ab},P_{c}\right] & = & i\left(\eta_{bc}P_{a}-\eta_{ac}P_{b}\right),\nonumber \\
\left[M_{ab},Z_{cd}\right] & = & i\left(\eta_{ad}Z_{bc}+\eta_{bc}Z_{ad}-\eta_{ac}Z_{bd}-\eta_{bd}Z_{ac}\right),\nonumber \\
\left[P_{a},D\right] & = & iP_{a},\nonumber \\
\left[Z_{ab},D\right] & = & 2iZ_{ab},\label{eq: mw algebra}
\end{eqnarray}
where $\eta_{ab}$ is the Minkowski metric which has $diag\left(\eta_{ab}\right)=\left(+1,-1,-1,-1\right)$,
the constant $\lambda$ has the unit of $L^{-2}$ which will be related
to the cosmological constant where $L$ is considered as the unit
of length and the indices $a,b,...=0,...,3$. Here, $M_{ab}$ is the
Lorentz generator, $P_{a}$ is the translation generator, $D$ is
the dilatation generator, and $Z_{ab}$ is the anti-symmetric Maxwell
symmetry generator. In addition to the Weyl algebra, $\mathcal{MW}$
algebra contains six new additional tensorial generators $Z_{ab}$.

Let us start to construct the gauge theory of gravity based on $\mathcal{MW}$
algebra. We first introduce a one-form gauge field that possesses
the following form,

\begin{equation}
\mathcal{A}=e^{a}P_{a}+B^{ab}Z_{ab}+\chi D-\frac{1}{2}\omega^{ab}M_{ab},\label{eq: gauge field 1}
\end{equation}
where the coefficient fields $e^{a}\left(x\right)=e_{\mu}^{a}\left(x\right)dx^{\mu}$,
$B^{ab}\left(x\right)=B_{\mu}^{ab}\left(x\right)dx^{\mu}$, $\chi\left(x\right)=\chi_{\mu}\left(x\right)dx^{\mu}$
and $\omega^{ab}\left(x\right)=\omega_{\mu}^{ab}\left(x\right)dx^{\mu}$
are the one-form gauge fields for the corresponding generators, respectively.
The transformation of these fields under infinitesimal gauge transformations
is given by $\delta\mathcal{A}=-d\zeta-i\left[\mathcal{A},\zeta\right]$
with the gauge generator,
\begin{equation}
\zeta\left(x\right)=y^{a}\left(x\right)P_{a}+\varphi^{ab}\left(x\right)Z_{ab}+\sigma\left(x\right)D-\frac{1}{2}\tau^{ab}\left(x\right)M_{ab},
\end{equation}
where $y^{a}\left(x\right)$ are space-time translations, $\varphi^{ab}\left(x\right)$
are translations in tensorial space, $\sigma\left(x\right)$ dilatation
parameter and $\tau^{ab}\left(x\right)$ are the Lorentz transformation
parameters. Thus we get,
\begin{eqnarray}
\delta e^{a} & = & -dy^{a}-\omega_{\,\,b}^{a}y^{b}-\chi y^{a}+\sigma e^{a}+\tau_{\,\,b}^{a}e^{b},\nonumber \\
\delta B^{ab} & = & -d\varphi^{ab}-\omega_{\,\,\,c}^{[a}\varphi^{c|b]}-2\chi\varphi^{ab}+\frac{1}{2}e^{[a}y^{b]}+2\sigma B^{ab}+\tau_{\,\,\,c}^{[a}B^{c|b]},\nonumber \\
\delta\chi & = & -d\sigma,\nonumber \\
\delta\omega^{ab} & = & -d\tau^{ab}-\omega_{\,\,\,c}^{[a}\tau^{c|b]},\label{delta_e}
\end{eqnarray}
where the antisymmetrization and symmetrization of the objects are
defined by $A_{[a}B_{b]}=A_{a}B_{b}-A_{b}B_{a}$ and $A_{(a}B_{b)}=A_{a}B_{b}+A_{b}B_{a}$,
respectively.

Using the structure equation $\mathcal{F}=d\mathcal{A}+\frac{i}{2}\left[\mathcal{A},\mathcal{A}\right]$
and defining the curvatures as $\mathcal{F}=\mathcal{F}^{a}P_{a}+\mathcal{F}^{ab}Z_{ab}+fD-\frac{1}{2}\mathcal{R}^{ab}M_{ab}$,
we find the associated two-form curvatures as,

\begin{eqnarray}
\mathcal{F}^{a} & = & de^{a}+\omega_{\,\,b}^{a}\wedge e^{b}+\chi\wedge e^{a},\nonumber \\
\mathcal{F}^{ab} & = & dB^{ab}+\omega_{\,\,\,c}^{[a}\wedge B^{c|b]}+2\chi\wedge B^{ab}-\frac{\lambda}{2}e^{a}\wedge e^{b},\nonumber \\
f & = & d\chi,\nonumber \\
\mathcal{R}^{ab} & = & d\omega^{ab}+\omega_{\,\,c}^{a}\wedge\omega^{cb}.\label{eq: curvatures}
\end{eqnarray}
The infinitesimal gauge transformation of the curvatures can be found
by the help of the equation $\delta\mathcal{F}=i\left[\zeta,\mathcal{F}\right]$
as follows

\begin{eqnarray}
\delta\mathcal{F}^{a} & = & \tau_{\,\,b}^{a}\mathcal{F}^{b}+\sigma\mathcal{F}^{a}-\mathcal{R}_{\,\,b}^{a}y^{b}-fy^{a},\nonumber \\
\delta\mathcal{F}^{ab} & = & \tau_{\,\,\,c}^{[a}\mathcal{F}^{c|b]}+2\sigma\mathcal{F}^{ab}+\frac{\lambda}{2}\mathcal{F}^{[a}y^{b]}-\mathcal{R}_{\,\,\,c}^{[a}\varphi^{c|b]}-2f\varphi^{ab},\nonumber \\
\delta f & = & 0,\nonumber \\
\delta\mathcal{R}^{ab} & = & \tau_{\,\,\,c}^{[a}\mathcal{R}^{c|b]}.
\end{eqnarray}
By performing the covariant derivative of the given curvatures, one
can derive the corresponding Bianchi identities in the following manner
\begin{eqnarray}
\mathcal{D}\mathcal{F}^{a} & = & \mathcal{R}_{\,\,b}^{a}\wedge e^{b}+f\wedge e^{a},\nonumber \\
\mathcal{D}\mathcal{F}^{ab} & = & \mathcal{R}_{\,\,\,c}^{[a}\wedge B^{c|b]}+2f\wedge B^{ab}-\frac{\lambda}{2}\mathcal{F}^{[a}\wedge e^{b]},\nonumber \\
\mathcal{D}f & = & 0,\nonumber \\
\mathcal{D}\mathcal{R}^{ab} & = & 0,\label{eq: Bianchi}
\end{eqnarray}
where $\mathcal{D}\Phi=[d+\omega+w(\Phi)\chi]$$\Phi$ is the Weyl
gauge covariant derivative and $w$ being the Weyl weight of the corresponding
field.

\subsection{Geometrical interpretation of Weyl's gauge theory}

In our scenario, constructing locally invariant Lagrangians it is
necessary to consider the scale invariance and its connection to the
dilatation subgroup of $\mathcal{MW}$. Achieving localization of
the dilatation symmetry leads us back to the Weyl gauge theory (for
more details, see \citep{Rosen:1982Weyl,Blagojevic:2002Gravitation,Karananas:2016WeylandRicciGauge,Hobson:2020WeylGauge}).
For this reason, we want to give a brief discussion of the geometrical
interpretation of the Weyl gauge theory.

The Weyl space and the Riemann space are two different mathematical
structures used to describe the geometry of space-time in the context
of general relativity and related theories. One of the key differences
is the conformal invariance. The Weyl space has conformal invariance,
which means that its geometry is invariant under changes in scale.
In contrast, the Riemann space does not have conformal invariance
and its geometry is not invariant under scale transformations.

The main point of the geometric interpretation is the identification
of $e_{a}^{\mu}\left(x\right)$ as the components of a vierbein system
in the Weyl space-time. This identification implies that the components
of the space-time metric tensor and the vierbein field are related
by the requirement that

\begin{equation}
g_{\mu\nu}=\eta_{ab}e_{\mu}^{a}e_{\nu}^{b},\label{eq: Metric_space-time}
\end{equation}
where $\mu,\nu=0,...,3$ are space-time indices and $\eta_{ab}$ is
the Minkowski metric which satisfies the following relation,
\begin{equation}
\eta_{ab}=g_{\mu\nu}e_{a}^{\mu}e_{b}^{\nu}.\label{eq: Metric_tangent}
\end{equation}

If one utilizes the gauge transformation for $e^{a}\left(x\right)$
as given in Eq.(\ref{delta_e}), it can be demonstrated that the transformation
of the space-time metric tensor satisfies the subsequent expression,
\begin{equation}
\delta g_{\mu\nu}\left(x\right)=2\sigma\left(x\right)g_{\mu\nu}\left(x\right),
\end{equation}
and the Minkowski metric satisfies $\delta\eta_{ab}=0$ under infinitesimal
gauge transformations. Moreover, it is known that the infinitesimal
scale transformations which act on a local generic field $\Phi\left(x\right)$
can be expressed as,

\begin{equation}
\delta\Phi\left(x\right)=w\sigma\left(x\right)\Phi\left(x\right),
\end{equation}
where $\sigma\left(x\right)$ is local parameter for the scale transformation
and $w$ is the ``Weyl weight'' (or simply \textquotedblleft weight\textquotedblright )
of scalar field $\Phi\left(x\right)$ (for more detail about the Weyl
transformation see \citep{Omote:1971Scale,Bregman:1973Weyl,Charap:1973GaugeWeyl,Kasuya:1975Weyl,Blagojevic:2002Gravitation,Babourova:2006Poincare-Weyl}).
Thus, the space-time metric and vierbein field have the Weyl weights
$w\left(g_{\mu\nu}\right)=2$, $w\left(g^{\mu\nu}\right)=-2$, $w\left(\sqrt{-g}\right)=4$,
and $w\left(e_{\mu}^{a}\right)=1$, respectively, and so (\ref{eq: Metric_space-time})
and (\ref{eq: Metric_tangent}) imply that $w\left(\eta_{ab}\right)=0$,
as expected.

In this geometry, the affine connection in the Weyl-Cartan space-time
is defined by

\begin{equation}
\Gamma_{\,\,\mu\nu}^{*\lambda}=\tilde{\Gamma}_{\,\,\mu\nu}^{\lambda}+\tilde{K}_{\,\,\mu\nu}^{\lambda},\label{eq: affine_connection}
\end{equation}
where $\tilde{\Gamma}_{\,\,\mu\nu}^{\lambda}=\tilde{\Gamma}_{\,\,\nu\mu}^{\lambda}$
is the Weyl connection which reads 
\begin{eqnarray}
\tilde{\Gamma}_{\,\mu\nu}^{\lambda} & = & \frac{1}{2}g^{\lambda\rho}\left(\tilde{\partial}_{\mu}g_{\nu\rho}+\tilde{\partial}_{\nu}g_{\mu\rho}-\tilde{\partial}_{\rho}g_{\mu\nu}\right)\nonumber \\
 & = & \Gamma_{\,\mu\nu}^{\lambda}+\delta_{\nu}^{\lambda}\chi_{\mu}+\delta_{\mu}^{\lambda}\chi_{\nu}-g_{\mu\nu}\chi^{\lambda},\label{eq: Weyl_connection}
\end{eqnarray}
where $\tilde{\partial}_{\mu}=\partial_{\mu}+w\chi_{\mu}$ and $\Gamma_{\,\,\mu\nu}^{\lambda}=\frac{1}{2}g^{\lambda\rho}\left(\partial_{\mu}g_{\nu\rho}+\partial_{\nu}g_{\mu\rho}-\partial_{\rho}g_{\mu\nu}\right)$
is the well-known Lorentz connection (or the Riemann-Christoffel connection).
The term $\tilde{K}_{\,\mu\nu}^{\lambda}$ in (\ref{eq: affine_connection})
is the Weyl-Cartan contortion tensor which has the antisymmetry property
in its first two indices as $\tilde{K}_{\lambda\mu\nu}=-\tilde{K}_{\mu\lambda\nu}^{\lambda}$
and it can be given in terms of the Weyl-Cartan torsion $\tilde{T}_{\,\mu\nu}^{\lambda}=\Gamma_{\,\,\left[\nu\mu\right]}^{*\lambda}$
by
\begin{equation}
\tilde{K}_{\,\mu\nu}^{\lambda}=-\frac{1}{2}\left(\tilde{T}_{\,\mu\nu}^{\lambda}-\tilde{T}_{\mu\,\,\nu}^{\,\,\lambda}+\tilde{T}_{\mu\nu}^{\,\,\,\lambda}\right).\label{eq: contorsion_tensor}
\end{equation}
It is clear from (\ref{eq: affine_connection})--(\ref{eq: contorsion_tensor})
that setting the torsion to zero ($\tilde{T}_{\,\mu\nu}^{\lambda}=0$)
then the Weyl-Cartan space-time reduces to the Weyl space-time. Together
with this condition, if dilatation gauge field $\chi_{\mu}\rightarrow0$,
then $\tilde{\Gamma}_{\,\mu\nu}^{\lambda}\rightarrow\Gamma_{\,\mu\nu}^{\lambda}$
and the Weyl geometry becomes Riemannian and the whole geometry is
completely defined by the metric tensor $g_{\mu\nu}$$\left(x\right)$.

Within the context of Weyl space-time, wherein torsion is absent,
the covariant differentiation of a vector $A^{\nu}$ formed with Weyl
connection may be expressed in the following manner,
\begin{eqnarray}
\tilde{\nabla}_{\mu}A^{\alpha} & = & \partial_{\mu}A^{\alpha}+\tilde{\Gamma}_{\,\,\mu\beta}^{\alpha}A^{\beta}\nonumber \\
 & = & \nabla_{\mu}A^{\alpha}+\delta_{\mu}^{\alpha}\chi_{\beta}A^{\beta}+A^{\alpha}\chi_{\mu}-A_{\mu}\chi^{\alpha},\label{eq: cov_der_Weyl}
\end{eqnarray}
where $\nabla_{\mu}A^{\alpha}=\partial_{\mu}A^{\alpha}+\Gamma_{\,\,\mu\beta}^{\alpha}A^{\beta}$
is the covariant derivative for diffeomorphisms in the Riemann geometry.
With the help of Eq.(\ref{eq: cov_der_Weyl}), the covariant derivative
of the metric tensor is conjectured to be nonzero
\begin{equation}
\tilde{\nabla}_{\rho}g_{\mu\nu}=-2\chi_{\rho}g_{\mu\nu}=Q_{\rho\mu\nu},\label{eq: semi-metricity}
\end{equation}
where the new geometric quantity $Q_{\rho\mu\nu}$ is called the nonmetricity
and Eq.(\ref{eq: semi-metricity}) is also called ``semi-metricity''
condition. Moreover, if we make use of the total covariant derivative
applied to the vierbein field $e_{\mu}^{a}\left(x\right)$, we get

\begin{equation}
\mathcal{D}_{\mu}\left(w+\tilde{\Gamma}\right)e_{\nu}^{a}=\tilde{\partial}_{\mu}e_{\nu}^{a}-\tilde{\Gamma}_{\nu\mu}^{\sigma}e_{\sigma}^{a}+\omega_{\,\,b\mu}^{a}e_{\nu}^{b}=0,\label{eq: cov_der_total}
\end{equation}
in which this relation is sometimes known as the \textquotedblleft tetrad
postulate\textquotedblright{} and demonstrates a connection between
$\omega_{\,\,b\mu}^{a}$, $\tilde{\Gamma}_{\,\mu\nu}^{\lambda}$ and
$\chi_{\mu}$. Now, it is straightforward to show that $\omega_{\,\,b\mu}^{a}$
or $\tilde{\Gamma}_{\,\mu\nu}^{\lambda}$ may be written explicitly
in terms of the other by using Eq.(\ref{eq: cov_der_total}) as follows,
\begin{equation}
\tilde{\Gamma}_{\mu\nu}^{\lambda}=e_{a}^{\lambda}\left(\tilde{\partial}_{\mu}e_{\nu}^{a}+\omega_{\,\,b\mu}^{a}e_{\nu}^{b}\right),\label{eq: affine_connection-1}
\end{equation}
and
\begin{equation}
\omega_{\,\,b\mu}^{a}=e_{\lambda}^{a}\left(\tilde{\partial}_{\mu}e_{b}^{\lambda}+\tilde{\Gamma}_{\nu\mu}^{\lambda}e_{b}^{\nu}\right),\label{eq: spin_connection}
\end{equation}
where we used $e_{\lambda}^{a}\tilde{\partial}_{\mu}e_{c}^{\lambda}=-e_{c}^{\lambda}\tilde{\partial}_{\mu}e_{\lambda}^{a}$
because $\tilde{\partial}_{\mu}\left(e_{\lambda}^{a}e_{c}^{\lambda}\right)=\tilde{\partial}_{\mu}\left(\delta_{c}^{a}\right)=0$.
If we substitute (\ref{eq: Weyl_connection}) in (\ref{eq: spin_connection}),
\begin{eqnarray}
\omega_{\,\,b\mu}^{a} & = & \mathring{\omega}_{\,\,b\mu}^{a}+\left(e_{\mu}^{a}e_{b}^{\nu}-e^{a\nu}e_{b\mu}\right)\chi_{\nu},\label{eq: spin_connection-2}
\end{eqnarray}
we get the Weyl spin connection in terms of $e_{\mu}^{a}$, $\chi_{\nu}$
and $\mathring{\omega}_{\,\,b\mu}^{a}$. Here $\mathring{\omega}_{\,\,b\mu}^{a}$
represents the standard spin connection for a torsionless theory which
is defined as
\begin{equation}
\mathring{\omega}_{\,\,b\mu}^{a}=e_{\lambda}^{a}\left(\partial_{\mu}e_{b}^{\lambda}+\Gamma_{\nu\mu}^{\lambda}e_{b}^{\nu}\right),\label{eq: spin_connection_Lorentz}
\end{equation}
or by using the definition of the Lorentz connection $\Gamma_{\nu\mu}^{\lambda}$,
it can be also written as
\begin{equation}
\mathring{\omega}_{\mu}^{ab}=\frac{1}{2}\left[e^{\nu a}\left(\partial_{\mu}e_{\nu}^{b}-\partial_{\nu}e_{\mu}^{b}\right)-e^{\nu b}\left(\partial_{\mu}e_{\nu}^{a}-\partial_{\nu}e_{\mu}^{a}\right)-e_{\mu c}e^{\nu a}e^{\lambda b}\left(\partial_{\nu}e_{\lambda}^{c}-\partial_{\lambda}e_{\nu}^{c}\right)\right].
\end{equation}
Subsequently, by utilizing Eq.(\ref{eq: spin_connection-2}) and (\ref{eq: spin_connection_Lorentz}),
one can express the gauge covariant derivative for a vector $A^{a}$
of weight $w$ as follows.

\begin{eqnarray}
\mathcal{D}_{\mu}A^{a} & = & \partial_{\mu}A^{a}+w\chi_{\mu}A^{a}+\omega_{\,\,b\mu}^{a}A^{b}\nonumber \\
 & = & \partial_{\mu}A^{a}+w\chi_{\mu}A^{a}+e_{\lambda}^{a}\left(\partial_{\mu}e_{b}^{\lambda}+\Gamma_{\nu\mu}^{\lambda}e_{b}^{\nu}\right)A^{b}+\left(e_{\mu}^{a}e_{b}^{\nu}-e^{a\nu}e_{b\mu}\right)\chi_{\nu}A^{b},
\end{eqnarray}
thus, we have formulated the covariant derivative in terms of $e_{\mu}^{a}$,
$\Gamma_{\,\mu\nu}^{\lambda}$ and $\chi_{\mu}$.

From this theoretical background, just as in the Riemannian geometry,
we can use the Weyl connection to compute a Weyl invariant curvature
tensor as follows
\begin{eqnarray}
\mathcal{R}_{\mu\nu\rho}{}^{\sigma} & = & \partial_{\nu}\tilde{\Gamma}_{\mu\rho}^{\sigma}-\partial_{\mu}\tilde{\Gamma}_{\nu\rho}^{\sigma}+\tilde{\Gamma}_{\mu\rho}^{\alpha}\tilde{\Gamma}_{\alpha\nu}^{\sigma}-\tilde{\Gamma}_{\nu\rho}^{\alpha}\tilde{\Gamma}_{\alpha\mu}^{\sigma}\nonumber \\
 & = & R_{\mu\nu\rho}{}^{\sigma}+\left(\delta_{[\mu}^{\sigma}\nabla_{\nu]}\chi_{\rho}-g_{[\mu\rho}\nabla_{\nu]}\chi^{\sigma}-\delta_{\rho}^{\sigma}f_{\mu\nu}\right)\nonumber \\
 &  & +\left(g_{\rho[\mu}\chi_{\nu]}\chi^{\sigma}-\delta_{[\mu}^{\sigma}\chi_{\nu]}\chi_{\rho}+\delta_{[\mu}^{\sigma}g_{\nu]\rho}\chi^{\alpha}\chi_{\alpha}\right),\label{eq: Weyl_curvature}
\end{eqnarray}
where $R_{\mu\nu\rho}{}^{\sigma}$ is the usual Riemann tensor and
$f_{\mu\nu}=\partial_{\mu}\chi_{\nu}-\partial_{\nu}\chi_{\mu}$ is
the Weyl invariant field strength. From this equation, it is possible
to establish a Weyl invariant Ricci tensor as,

\begin{equation}
\mathcal{R}_{\mu\rho}=\mathcal{R}_{\mu\sigma\rho}{}^{\sigma}=R_{\mu\rho}-\left(g_{\mu\rho}\nabla_{\sigma}\chi^{\sigma}+2\nabla_{\mu}\chi_{\rho}+f_{\mu\rho}\right)+2\left(\chi_{\mu}\chi_{\rho}-g_{\mu\rho}\chi^{\alpha}\chi_{\alpha}\right),\label{eq: Ricci_tensor_weyl}
\end{equation}
where $R_{\mu\rho}$ is the Ricci tensor for the Riemannian geometry.
Finally, one finds that the scalar curvatures $\mathcal{R}$ of Weyl
geometry and $R$ of the Riemannian geometry are related by
\begin{equation}
\mathcal{R}=g^{\mu\rho}\mathcal{R}_{\mu\rho}=R-6\nabla_{\mu}\chi^{\mu}-6\chi_{\mu}\chi^{\mu}.\label{eq: Ricci_scalar_weyl}
\end{equation}

\subsection{Gravitational action}

The formulation of a Lagrangian based on $\mathcal{MW}$ algebra differs
somewhat from the ordinary Maxwell algebra because $\mathcal{MW}$
algebra contains the scale transformations. In the gauge theory of
gravity based on $\mathcal{MW}$ algebra, the well-known action $S=\int d^{4}x\sqrt{-g}R$
is not invariant under the local scale transformations. To overcome
this difficulty, Weyl introduced quadratic terms into the action to
establish a consistent theory \citep{Weyl:1922SpaceTimeMatter}. Furthermore,
Brans and Dicke constructed a Lagrangian with the combination of the
scalar curvature $\mathcal{R}$ and a new compensating scalar field
$\phi\left(x\right)$ \citep{Brans:1961sx}. Subsequently, Dirac employed
a scalar field $\phi\left(x\right)$ with a Weyl weight of $-1$ and
constructed the action using $\phi^{2}\mathcal{R}$ rather than $\mathcal{R}$
\citep{Dirac:1973LongRange} (for more information, see \citep{Agnese:1975GaugeFields,Dereli:1982WEYL}).

These theoretical foundations were used to introduce the gravitational
action in the Maxwell-Weyl gauge theory in \citep{cebeciouglu2014gauge},
following Dirac's convention. In this study, a scalar field $\phi\left(x\right)$
(Dirac scalar field) is introduced which transforms as $\delta\phi=-\sigma\phi$
to make the Einstein-Hilbert action to be local scale invariant (for
more detail, see \citep{Dirac:1973LongRange,cebeciouglu2014gauge}).

In the present context, it is possible to introduce a shifted curvature
of zero Weyl weight as,
\begin{equation}
\mathcal{J}^{ab}=\mathcal{R}^{ab}+2\mu\phi^{2}\mathcal{F}^{ab},
\end{equation}
where $\mu$ is an arbitrary constant. Using this definition, one
can express a gravitational action with zero Weyl weight in the following
manner \citep{cebeciouglu2014gauge},

\begin{eqnarray}
S & = & \int\frac{1}{4\kappa\lambda\mu}\mathcal{J}\wedge^{\ast}\mathcal{J}+\frac{1}{4}f\wedge^{\ast}f-\frac{1}{2}\mathcal{D}\phi\wedge^{\ast}\mathcal{D}\phi+\frac{\varsigma}{4}\phi^{4}{}^{\ast}1,\label{eq:lagrangian_euler-1}
\end{eqnarray}
where $\kappa=8\pi G$ and $\varsigma$ are constants and $*$ represents
the Hodge duality operator which is used as $\mathcal{J}\wedge^{\ast}\mathcal{J}=\frac{1}{2}\varepsilon_{abcd}\mathcal{J}^{ab}\wedge\mathcal{J}^{cd}$.
The first term represents the free gravitational part, the second
term is the Maxwell-like kinetic term and the last two terms correspond
to the kinetic term and the self-interacting term for the Dirac scalar
field $\phi\left(x\right)$.

The equations of motion can be found by taking variations of Eq.(\ref{eq:lagrangian_euler-1})
with respect to gauge fields $\omega^{ab}\left(x\right)$, $e^{a}\left(x\right)$,
$B^{ab}\left(x\right)$, $\chi\left(x\right)$, and $\phi\left(x\right)$,
respectively,
\begin{equation}
\mathcal{D}\mathcal{J}^{ab}-2\mu\phi^{2}\mathcal{J}_{\,\,\,\,e}^{[a}\wedge B^{e|b]}=0,\label{eq: omega_var}
\end{equation}

\begin{eqnarray}
 &  & -\frac{\phi^{2}}{\kappa}\varepsilon_{abcd}\mathcal{J}^{ab}\wedge e^{d}+\frac{1}{2}[\mathcal{D}_{c}\phi^{\ast}\mathcal{D}\phi+\mathcal{D}\phi\wedge^{\ast}\left(e_{a}\wedge e_{c}\right)\mathcal{D}^{a}\phi]\nonumber \\
 &  & -\frac{1}{2}(f_{cb}e^{b}\wedge^{\ast}f-\frac{1}{2}\varepsilon_{abcd}f^{ab}e^{d}\wedge f)+\frac{\varsigma\phi^{4}}{2}{}^{\ast}e_{c}=0,\label{e_var}
\end{eqnarray}

\begin{equation}
\mathcal{D}(\phi^{2}\mathcal{J}^{ab})=0,\label{eq: B_var}
\end{equation}

\begin{equation}
\frac{\phi^{2}}{\kappa\lambda}\epsilon_{abcd}\mathcal{J}^{ab}\wedge B^{cd}+\frac{1}{2}\mathcal{D}^{\ast}f+\phi^{\ast}\mathcal{D}\phi=0,\label{eq: chi_var}
\end{equation}

\begin{equation}
\frac{\phi}{\kappa\lambda}\epsilon_{abcd}\mathcal{J}^{ab}\wedge\mathcal{F}^{cd}+\mathcal{D}^{\ast}\mathcal{D}\phi+\varsigma\phi^{3}{}^{\ast}1=0.\label{eq: phi_var}
\end{equation}
These equations are invariant with respect to the local Maxwell-Weyl
transformation considered above. By a straightforward calculation,
passing from tangent space to world indices for Eq.(\ref{e_var}),
one gets the following field equation with cosmological term depending
on the dilaton field,
\begin{equation}
\mathcal{R}_{\alpha}^{\mu}-\frac{1}{2}\delta_{\alpha}^{\mu}\mathcal{R}+3\mu\lambda\phi^{2}\delta_{\alpha}^{\mu}=-2\mu\phi^{2}T_{B}{}_{\,\,\alpha}^{\mu}-\kappa\phi^{-2}\left(T_{\phi}{}_{\,\,\alpha}^{\mu}+\frac{1}{2}T_{f}{}_{\,\,\alpha}^{\mu}\right),\label{eq: EFE}
\end{equation}
where 
\begin{eqnarray}
T_{B}{}_{\,\,\alpha}^{\mu} & = & e_{a}^{\mu}e_{b}^{\beta}\mathcal{D}_{[\alpha}B_{\beta]}^{ab}-\frac{1}{2}\delta_{\alpha}^{\mu}\left(e_{a}^{\rho}e_{b}^{\sigma}\mathcal{D}_{[\rho}B_{\sigma]}^{ab}\right),\nonumber \\
T_{\phi}{}_{\,\,\alpha}^{\mu} & = & \mathcal{D}^{\mu}\phi\mathcal{D}_{\alpha}\phi-\frac{1}{2}\delta_{\alpha}^{\mu}\left(\mathcal{D}_{\gamma}\phi\mathcal{D}^{\gamma}\phi-\frac{\varsigma}{2}\phi^{4}\right),\nonumber \\
T_{f}{}_{\,\,\alpha}^{\mu} & = & f^{\mu\beta}f_{\beta\alpha}+\frac{1}{4}\delta_{\alpha}^{\mu}f_{\gamma\delta}f^{\gamma\delta},\label{eq: em_tensors_1}
\end{eqnarray}
are the energy-momentum tensors for the $B$-gauge field, Dirac scalar
field and dilatation curvature, respectively. Note that one can unify
the energy-momentum tensor $T_{B}{}_{\,\,\alpha}^{\mu}$ and the term
$3\mu\lambda\phi^{2}\delta_{\alpha}^{\mu}$ because both of these
terms originated from the Maxwell curvature. But we prefer to use
these quantities separately to show the cosmological constant explicitly.

Let us examine the equation of motion with respect to $\omega^{ab}\left(x\right)$
in Eq.(\ref{eq: omega_var}). In this equation, we see that $\mathcal{R}^{ab}=2\mu\phi^{2}\mathcal{F}^{ab}$
condition arises as a special solution. Then as a consequence of this
condition, Eq.(\ref{eq: chi_var}) reduces its original form which
is derived in the Weyl gauge theory \citep{Rosen:1982Weyl}. By the
help of this condition, we introduce the following gauge which was
used in Rosen's paper \citep{Rosen:1982Weyl} as ``\textit{the cosmic
gauge}'' derived from equation of motion with respect to $\chi\left(x\right)$
in Eq.(\ref{eq: chi_var}) to find cosmological solutions for Weyl's
gauge theory;
\begin{equation}
\mathcal{D}_{\mu}\phi=\partial_{\mu}\phi-\chi_{\mu}\phi=0.\label{eq: cosmic_gauge}
\end{equation}
By using this gauge, we can establish the dilatation gauge field by
expressing it as the gradient of the Dirac scalar field in the following
manner
\begin{equation}
\chi_{\mu}=\partial_{\mu}\ln\phi.\label{eq: constraint}
\end{equation}
Therefore this constraint leads to the dilatation curvature $f_{\mu\nu}=\partial_{\mu}\chi_{\nu}-\partial_{\nu}\chi_{\mu}$
in Eq.(\ref{eq: curvatures}) to be zero. This is another way of expressing
the inverse Higgs effect \citep{Ivanov:1975TheInverseHiggs,Low:2002SpontaneouslyBroken}.
Clearly, this also shows the equivalence of the inverse Higgs constraint
to equations of motion. Furthermore, as a result of these constraints,
the constant $\varsigma$ and the energy-momentum tensors $T_{\phi}{}_{\,\,\alpha}^{\mu}$
and $T_{f}{}_{\,\,\alpha}^{\mu}$ in Eq.(\ref{eq: em_tensors_1})
should be vanish. It is worth noting that one can show that the equation
of motions in Eqs.(\ref{eq: omega_var})-(\ref{eq: phi_var}) satisfy
each other under these constraints.

Given this context, considering the cosmic gauge in Eq.(\ref{eq: cosmic_gauge}),
the field equation in Eq.(\ref{eq: B_var}) and the Bianchi identity
$\mathcal{D}\mathcal{R}^{ab}=0$ in Eq.(\ref{eq: Bianchi}), it can
be demonstrated that $\mathcal{D}\mathcal{F}^{ab}=0$. This result
establishes the necessary background to work in the torsion-free condition
so we can use the Weyl invariant curvature tensor as given in Eqs.(\ref{eq: Weyl_curvature})-(\ref{eq: Ricci_scalar_weyl}).

Substituting Eqs.(\ref{eq: Ricci_tensor_weyl}) and (\ref{eq: Ricci_scalar_weyl})
into Eq.(\ref{eq: EFE}), the field equation is reduced to the following
form,

\begin{equation}
G_{\,\,\alpha}^{\mu}+\Lambda\phi^{2}\delta_{\,\,\alpha}^{\mu}=T_{\chi}{}_{\,\,\alpha}^{\mu}-2\mu\phi^{2}T_{B}{}_{\,\,\alpha}^{\mu},\label{eq: EFE-1}
\end{equation}
where $G_{\,\,\alpha}^{\mu}=R_{\,\,\alpha}^{\mu}-\frac{1}{2}\delta_{\,\,\alpha}^{\mu}R$
is the well-known Einstein tensor and 
\begin{equation}
T_{\chi}{}_{\,\,\alpha}^{\mu}=2\left[\nabla^{\mu}\chi_{\alpha}-\chi^{\mu}\chi_{\alpha}+\frac{1}{2}g^{\mu\beta}\left(\partial_{\beta}\chi_{\alpha}-\partial_{\alpha}\chi_{\beta}\right)\right]-\delta_{\alpha}^{\mu}\left(2\nabla_{\sigma}\chi^{\sigma}+\chi_{\sigma}\chi^{\sigma}\right).
\end{equation}
is the energy-momentum tensor with respect to the dilatation gauge
field. Furthermore, it is worth noting that the term $\Lambda\phi^{2}$
can be seen as the effective cosmological constant and therefore this
term can also be interpreted as the effective dark energy density
\citep{Babourova:2012Gravitationtheory}. Here, the constant $\Lambda$
is given by

\begin{equation}
\Lambda=3\mu\lambda.\label{eq:cosm_cons}
\end{equation}

On the other hand, in the presence of matter the action which is invariant
under the Weyl transformation, the gravitational field is given by
$S=S_{g}+S_{M}$ similar to \citep{Rosen:1982Weyl}. Thus, we can
write the field equation as follows,
\begin{equation}
G_{\,\,\alpha}^{\mu}+\Lambda\phi^{2}\delta_{\,\,\alpha}^{\mu}=\frac{\kappa}{\phi^{2}}T_{M}{}_{\,\,\alpha}^{\mu}+T_{\chi}{}_{\,\,\alpha}^{\mu}-2\mu\phi^{2}T_{B}{}_{\,\,\alpha}^{\mu},\label{eq: EFE-3}
\end{equation}
where $T_{M}{}_{\,\,\alpha}^{\mu}$ is the ordinary matter energy-momentum
tensor which is covariantly conserved, $\nabla_{\mu}T_{M}{}_{\,\,\alpha}^{\mu}=0$.

At this point, we assume that the matter-energy constituents of the
Universe are governed by a perfect fluid, which may be characterized
by two thermodynamic variables: the energy density $\rho\left(t\right)$
and the pressure $P\left(t\right)$. Accordingly, the complete energy-momentum
tensor generated by the Maxwell-Weyl gauge theory of gravity takes
the following form:
\begin{equation}
T_{\,\,\alpha}^{\mu}=diag\left(\rho,-P,-P,-P\right),
\end{equation}
which can be defined as follows by using Eq.(\ref{eq: EFE-3}),
\begin{eqnarray}
T_{\,\,\alpha}^{\mu} & = & \frac{\kappa}{\phi^{2}}T_{M}{}_{\,\,\alpha}^{\mu}+T_{\chi}{}_{\,\,\alpha}^{\mu}-2\mu\phi^{2}T_{B}{}_{\,\,\alpha}^{\mu}+T_{\Lambda}{}_{\,\,\alpha}^{\mu}.
\end{eqnarray}
Thus, Eq.(\ref{eq: EFE-3}) can be written in a compact form as $G_{\alpha}^{\mu}=T{}_{\,\,\alpha}^{\mu}$.
In this case, the total energy density and total pressure are defined
by the combination of the matter, dilatation, Maxwell gauge field
and the effective cosmological constant contributions as follows
\begin{equation}
\rho=\rho_{M}+\rho_{\chi}+\rho_{B}+\rho_{\Lambda},\label{eq: rho_total}
\end{equation}
and
\begin{equation}
P=P_{M}+P_{\chi}+P_{B}+P_{\Lambda},\label{eq: P_total}
\end{equation}
where we assume that all energy densities are positive definite. The
conservation of the energy-stress tensor in the FLRW background is
represented by the equation $\nabla_{\mu}T^{\mu\nu}=0$, which consequently
gives rise to the continuity equation
\begin{equation}
\dot{\rho}+3H\left(\rho+P\right)=0,
\end{equation}
where $H\left(t\right)$ is the Hubble parameter.

\section{Modified Friedmann Equations\label{sec:Cosmological-setup}}

The Einstein-Yang-Mills cosmology is a theoretical model that combines
the principles of general relativity and Yang-Mills theory to study
the evolution of the universe as a whole. The model proposes that
the gravitational force can be described in terms of a connection
between space-time and a non-Abelian gauge field. This model employed
a non-abelian gauge field to study the behavior of the universe during
the early stages of its evolution, including the inflationary period
and the formation of cosmic structures such as galaxies and galaxy
clusters. Similar to this theory, instead of the non-abelian Yang-Mills
fields, we consider finding potential cosmological consequences of
the Maxwell gauge fields $B_{\mu}^{ab}$ which are present in the
gravitational field equation in Eq.(\ref{eq: EFE}).

In general, for most cosmological models, the metric is assumed to
be a flat FLRW metric which based on Einstein's equations of general
relativity, provides a framework for describing the evolution of the
universe by assuming homogeneity and isotropy on large scales. So
we start from the following line element,

\begin{eqnarray}
ds^{2} & = & dt^{2}-a\left(t\right)^{2}\left(dx^{2}+dy^{2}+dz^{2}\right),\label{eq: FLRW metric}
\end{eqnarray}
where $a\left(t\right)$ is the scale factor. In accordance with our
assumptions, let us take the Dirac scalar field as a function of the
cosmic time $\phi\left(x\right)\rightarrow\phi\left(t\right)$. Thus,
by using Eq.(\ref{eq: constraint}), the dilatation gauge field can
be identified with only one nonzero component;
\begin{equation}
\chi_{\mu}=\left(\frac{\dot{\phi}}{\phi},0,0,0\right).
\end{equation}
Moreover, according to \citep{Azcarraga2013maxwellApplication}, we
can define the Maxwell gauge fields $B_{\mu}^{ab}\left(x\right)$
in terms of one-dimensional time dependent fields $\psi\left(t\right)$
and $\zeta\left(t\right)$,

\begin{equation}
B_{\mu}^{0s}\left(x\right)\rightarrow B_{\mu}^{0s}\left(t\right)=\left(0,\delta_{i}^{s}\psi\left(t\right)\right),\label{eq: B_1}
\end{equation}

\begin{equation}
B_{\mu}^{rs}\left(x\right)\rightarrow B_{\mu}^{rs}\left(t\right)=\left(0,\epsilon_{i}^{rs}\zeta\left(t\right)\right),\label{eq: B_2}
\end{equation}
where $r,s=1,2,3$ are the tangent indices and $i,j=1,2,3$ are the
space-time indices. Here, $\delta_{i}^{s}$ and $\epsilon_{i}^{rs}$
are three dimensional $so\left(3\right)$ tensors.

Having established the necessary background, we can now proceed to
derive the Friedmann equations. By employing Eqs.(\ref{eq: B_1}),
(\ref{eq: B_2}) and (\ref{eq: spin_connection_Lorentz}) and subsequently
applying them to the field equation given in Eq.(\ref{eq: EFE}),
the following equation can be obtained by focusing on the $\left(0,0\right)$
component of Eq.(\ref{eq: EFE}),

\begin{equation}
\left(\frac{\dot{a}}{a}\right)^{2}=\frac{\kappa\rho_{M}}{3\phi^{2}}-\frac{2\dot{a}\dot{\phi}}{a\phi}-\left(\frac{\dot{\phi}}{\phi}\right)^{2}-\frac{4\mu\phi^{2}\psi}{a}\left(\frac{\dot{a}}{a}+\frac{\dot{\phi}}{\phi}\right)-\frac{\Lambda\phi^{2}}{3},\label{eq: Friedmann_00}
\end{equation}
where the dot denotes the derivative with respect to the cosmological
time $t$. The $\left(i,i\right)$ components of Eq.(\ref{eq: EFE})
lead to,
\begin{equation}
\frac{2\ddot{a}}{a}+\left(\frac{\dot{a}}{a}\right)^{2}=-\frac{\kappa P_{M}}{\phi^{2}}-\frac{2\ddot{\phi}}{\phi}-\frac{4\dot{a}\dot{\phi}}{a\phi}+\left(\frac{\dot{\phi}}{\phi}\right)^{2}-\frac{4\mu\phi^{2}\psi}{a}\left(\frac{\dot{a}}{a}+\frac{\dot{\psi}}{\psi}+5\frac{\dot{\phi}}{\phi}\right)-\Lambda\phi^{2}.\label{eq: Friedmann_ii}
\end{equation}
Then by making use of Eqs.(\ref{eq: Friedmann_00}) and (\ref{eq: Friedmann_ii}),
the acceleration equation can be obtained as,

\begin{equation}
\frac{\ddot{a}}{a}=-\frac{\kappa}{6\phi^{2}}\left(3P_{M}+\rho_{M}\right)-\frac{\dot{a}\dot{\phi}}{a\phi}+\left(\frac{\dot{\phi}}{\phi}\right)^{2}-\frac{\ddot{\phi}}{\phi}-\frac{2\mu\phi^{2}\psi}{a}\left(\frac{\dot{\psi}}{\psi}+\frac{4\dot{\phi}}{\phi}\right)-\frac{\Lambda\phi^{2}}{3},\label{eq: Friedmann_acceleration}
\end{equation}
which can also be obtained by considering the trace of Eq.(\ref{eq: EFE}).
For completeness, if we combine Eqs.(\ref{eq: Friedmann_00}) and
(\ref{eq: Friedmann_acceleration}), we also derive the time derivative
of the Hubble parameter ($H\left(t\right)=\dot{a}\left(t\right)/a\left(t\right)$)
as
\begin{eqnarray}
\dot{H} & = & -\frac{\kappa}{2\phi^{2}}\left(\rho_{M}+P_{M}\right)+\frac{\dot{a}\dot{\phi}}{a\phi}+2\left(\frac{\dot{\phi}}{\phi}\right)^{2}-\frac{\ddot{\phi}}{\phi}+\frac{2\mu\phi^{2}}{a^{2}}\left(2\dot{a}\psi-a\dot{\psi}-\frac{2a\dot{\phi}\psi}{\phi}\right)\nonumber \\
 & = & -\frac{\kappa}{2\phi^{2}}\left(\rho+P\right).\label{eq: H_der}
\end{eqnarray}

Furthermore, by using Eqs.(\ref{eq: Friedmann_00}) and (\ref{eq: Friedmann_ii}),
one can find the energy density and the pressure expressions for the
corresponding fields, as presented below;

\begin{equation}
\rho_{\chi}=-\frac{3\phi^{2}}{\kappa}\left[\frac{2\dot{a}\dot{\phi}}{a\phi}+\left(\frac{\dot{\phi}}{\phi}\right)^{2}\right],\,\,\,\,\,\,\,\,\,P_{\chi}=\frac{\phi^{2}}{\kappa}\left[\frac{4\dot{a}\dot{\phi}}{a\phi}-\left(\frac{\dot{\phi}}{\phi}\right)^{2}+\frac{2\ddot{\phi}}{\phi}\right],\label{eq: rho_P_chi}
\end{equation}

\begin{equation}
\rho_{\Lambda}=-\frac{\Lambda\phi^{4}}{\kappa},\,\,\,\,\,\,\,\,\,P_{\Lambda}=\frac{\Lambda\phi^{4}}{\kappa},\label{eq: rho_P_Lambda-m}
\end{equation}

\begin{eqnarray}
\rho_{B} & = & -\frac{12\mu\phi^{4}\psi}{\kappa a}\left(\frac{\dot{a}}{a}+\frac{\dot{\phi}}{\phi}\right),\,\,\,\,\,\,\,\,\,P_{B}=\frac{4\mu\phi^{4}\psi}{\kappa a}\left(\frac{\dot{a}}{a}+\frac{\dot{\psi}}{\psi}+5\frac{\dot{\phi}}{\phi}\right).\label{eq: rho_P_B-m}
\end{eqnarray}
Let's look at the conditions that make energy densities positive.
For the positive values of the energy density $\rho_{\chi}$ it is
required that $2\dot{a}/a<-\dot{\phi}/\phi$ condition should be satisfied.
Regarding $\rho_{\Lambda}$, $\Lambda$ should be negative so $\mu$
or $\lambda$ can be chosen as a negative constant. Finally, when
we look at $\rho_{B}$ we have two conditions under $\mu=-1$ selection:
For positive values of $\psi\left(t\right)$ function, we get $\dot{a}/a+\dot{\phi}/\phi>0$,
and for negative values of $\psi\left(t\right)$, we have $\dot{a}/a+\dot{\phi}/\phi<0$
condition.

Furthermore, Eqs.(\ref{eq: Friedmann_00})-(\ref{eq: Friedmann_acceleration})
represent the time evolution of the Universe. For instance, when we
look at Eq.(\ref{eq: Friedmann_acceleration}), this equation demonstrates
that there is a time dependent Universe expansion which is controlled
by two time dependent functions and two constants; $\phi\left(t\right)$,
$\psi\left(t\right)$, $\mu$ and $\lambda$ respectively in the matterless
case ($\rho_{M}=P_{M}=0$). If we consider $\mu=-1$, the acceleration
equation reduces to the following form
\begin{equation}
\frac{\ddot{a}}{a}=-\frac{\dot{a}\dot{\phi}}{a\phi}+\left(\frac{\dot{\phi}}{\phi}\right)^{2}+\lambda\phi^{2}-\frac{\ddot{\phi}}{\phi}+\frac{2\phi^{2}\psi}{a}\left(\frac{\dot{\psi}}{\psi}+\frac{4\dot{\phi}}{\phi}\right),
\end{equation}
where the first three terms give positive contribution to the acceleration
and the remaining terms may generate positive or negative contributions.
It can easily be seen from this equation that to achieve a positive
acceleration ($\ddot{a}>0$) solutions, the following condition should
be satisfied,
\begin{equation}
-\frac{\ddot{\phi}}{\phi}+\frac{2\phi^{2}\psi}{a}\left(\frac{\dot{\psi}}{\psi}+\frac{4\dot{\phi}}{\phi}\right)>\frac{\dot{a}\dot{\phi}}{a\phi}-\left(\frac{\dot{\phi}}{\phi}\right)^{2}-\lambda\phi^{2}.
\end{equation}

On the other hand, by setting $\psi\left(t\right)=0$, the equations
(\ref{eq: Friedmann_00}), (\ref{eq: Friedmann_ii}) and (\ref{eq: Friedmann_acceleration})
can be reduced to the forms presented in Rosen's paper \citep{Rosen:1982Weyl}.
Additionally, imposing the compensating field $\phi\left(t\right)=1$
yields the equations derived in \citep{Kibaroglu:2022CosmoMaxwell}
for the Maxwell algebra. Finally, the conventional cosmological equations
of general relativity, along with the cosmological term, can be obtained
by setting $\psi\left(t\right)=0$, $\phi\left(t\right)=1$ and $\mu=-1$
as follows,

\begin{eqnarray}
\left(\frac{\dot{a}}{a}\right)^{2} & = & \frac{\kappa\rho_{M}}{3}+\lambda,\nonumber \\
\frac{2\ddot{a}}{a}+\left(\frac{\dot{a}}{a}\right)^{2} & = & -\kappa P_{M}+3\lambda,\nonumber \\
\frac{\ddot{a}}{a} & = & -\frac{\kappa}{6}\left(\rho_{M}+3P_{M}\right)+\lambda,\label{eq:Friedmann_classical}
\end{eqnarray}
where the constant $\lambda$ which is the central charge that characterizes
the Maxwell extension (\ref{eq: mw algebra}) provide a geometrical
way to introduce the cosmological constant similar to the de Sitter
case where $\left[P_{a},P_{b}\right]=\frac{1}{R^{2}}M_{ab}$, $R$
is the de Sitter radius and the cosmological constant is identified
as $\Lambda=\frac{1}{R^{2}}$.

\section{Examining Various Cosmological Scenarios\label{sec:Applications-for-various}}

Let us proceed with the Friedmann equations in the absence of a matter
source. Solving Eq.(\ref{eq: Friedmann_acceleration}) for $\psi\left(t\right)$,
we obtain the following equation:

\begin{equation}
\psi\left(t\right)=\frac{1}{\phi^{4}}\left[\frac{1}{6\mu}\int\left(\Lambda\phi^{4}a+3\ddot{\phi}\phi a-3\dot{\phi}^{2}\dot{a}+3\dot{\phi}\phi\dot{a}+3\phi^{2}\ddot{a}\right)dt+C_{1}\right].\label{eq: psi_depend}
\end{equation}
Then substituting this equation into Eq.(\ref{eq: Friedmann_00})
and solve with respect to $\phi\left(t\right)$ we get
\begin{equation}
\phi\left(t\right)=\frac{C_{2}}{a\sqrt{\mathcal{A}}},\label{eq: phi_exact}
\end{equation}
where we have introduced a new expression for simplicity as follows
\begin{equation}
\mathcal{A}\left(t\right)=C_{1}+\int\frac{1}{a^{3}}dt.
\end{equation}
Here $C_{1}$, $C_{2}$ are integral constants. Eq.(\ref{eq: phi_exact})
shows the direct relationship between $a\left(t\right)$ and $\phi\left(t\right)$
and therefore Eq.(\ref{eq: psi_depend}) takes the following form
\begin{equation}
\psi\left(t\right)=-\frac{\mathcal{A}^{2}a^{4}}{C_{2}^{4}}\left[\frac{C_{2}^{2}}{12\mu}\int\left(\frac{2\mathcal{A}a^{2}\left(\Lambda C_{2}^{2}a^{2}+3\dot{a}\right)+3}{a^{7}\mathcal{A}^{3}}\right)dt-C_{1}\right].\label{eq: psi_exact}
\end{equation}
Finally, we have found an exact relationship between the function
$\psi\left(t\right)$ and the scale factor $a\left(t\right)$. Utilizing
the solutions in Eqs.(\ref{eq: phi_exact}) and (\ref{eq: psi_exact}),
we can examine how various cosmological scenarios can be realized
within the framework of this generalized gravitation model. Let's
explore these solutions alongside different cosmological scenarios
which provide unique insights into different phases of cosmic evolution,
from the rapid expansion of the early universe to potential cyclical
patterns of contraction and rebirth.

\subsection{Power-law cosmology}

Power law cosmological evolution refers to a class of solutions to
the Friedmann equations where the scale factor $a\left(t\right)$
evolves as a power of time like $a\left(t\right)\propto t^{n}$ where
$n$ is a constant that depends on the dominant component of the universe's
energy density. This type of evolution is characteristic of universes
dominated by different types of matter or energy, such as radiation
and matter. Solving the Friedmann equations for a radiation-dominated
universe yields: $a\left(t\right)\propto t^{1/2}$. This leads to
the equation of state (EoS) function $\omega=1/3$, where $\omega$
is the ratio of pressure to energy density. For a matter-dominated
universe, the solution to the Friedmann equations is: $a\left(t\right)\propto t^{2/3}$
($\omega=0$). This describes the expansion of the universe during
the matter-dominated era.

Let us analyze a general class of the power-law cosmology as
\begin{equation}
a\left(t\right)=a_{0}\left(\frac{t}{t_{0}}\right)^{n},
\end{equation}
where $t_{0}$ is present age of the universe and $a_{0}$ is the
present value of $a\left(t\right)$. In this case, the Dirac scalar
function $\phi\left(t\right)$ (\ref{eq: phi_exact}) and the scalar
function comes from the construction of the Maxwell gauge field $\psi\left(t\right)$
(\ref{eq: psi_exact}) transform into the following forms;

\begin{equation}
\phi\left(t\right)=\frac{C_{2}}{a_{0}\left(t/t_{0}\right)^{n}}\sqrt{\frac{3\left(n-1\right)a_{0}^{3}}{3a_{0}^{3}C_{1}k-t\left(t/t_{0}\right)^{-3n}}},
\end{equation}

\begin{eqnarray}
\psi\left(t\right) & = & \frac{1}{1944C_{1}C_{2}^{4}\mu a_{0}^{2}k^{2}\left[C_{1}a_{0}^{3}k\left(\frac{t}{t_{0}}\right)^{3n}-\frac{t}{3}\right]^{2}}\Biggl\{-2592C_{1}^{3}a_{0}^{9}k^{3}\left(\mu C_{1}^{2}-\frac{\Lambda C_{2}^{4}}{24}\right)t\left(\frac{t}{t_{0}}\right)^{7n}\nonumber \\
 &  & -162C_{1}^{2}C_{2}^{2}a_{0}^{5}k^{3}t\left(\frac{t}{t_{0}}\right)^{3n}+1296C_{1}^{2}a_{0}^{6}k^{2}\left(\mu C_{1}^{2}-\frac{\Lambda C_{2}^{4}}{12}\right)t^{2}\left(\frac{t}{t_{0}}\right)^{4n}\nonumber \\
 &  & +1944C_{1}^{6}\mu a_{0}^{12}k^{4}\left(\frac{t}{t_{0}}\right)^{10n}+243C_{1}^{3}C_{2}^{2}a_{0}^{8}k^{4}\left(\frac{t}{t_{0}}\right)^{6n}\nonumber \\
 &  & +27t^{2}\left[\frac{8}{9}\left(\mu C_{1}^{2}-\frac{\Lambda C_{2}^{4}}{6}\right)t^{2}\left(\frac{t}{t_{0}}\right)^{-2n}-C_{1}a_{0}^{2}k\left[\frac{32a_{0}}{3}\left(\mu C_{1}^{2}-\frac{\Lambda C_{2}^{4}}{8}\right)t\left(\frac{t}{t_{0}}\right)^{n}-C_{2}^{2}k\right]\right]\Biggr\},
\end{eqnarray}
where we have introduced $k=n-1/3$ for simplicity. 

\subsection{de Sitter like evolution}

The de Sitter universe is a solution to Einstein's field equations
and this model is significant in the study of cosmological evolution,
especially in the context of dark energy and the accelerated expansion
of the universe. The de Sitter metric is often written in the form:

\begin{equation}
ds^{2}=-dt^{2}+e^{2H_{0}t}\left(dx^{2}+dy^{2}+dz^{2}\right),
\end{equation}
where $H_{0}$ is the Hubble parameter and is a constant. In this
universe, the scale factor grows exponentially with time as $a\left(t\right)=e^{H_{0}t}$
indicating that the universe undergoes accelerated expansion. This
is in contrast to power-law expansion in matter or radiation-dominated
universes. The de Sitter space is characterized by a vacuum-dominated
energy density, meaning that the energy density is constant and the
pressure is negative, $P=-\rho$ ($\omega=-1$). 

Therefore, the scalar functions in (\ref{eq: phi_exact}) and (\ref{eq: psi_exact})
become

\begin{equation}
\phi\left(t\right)=\frac{C_{\boldsymbol{2}}}{e^{H_{0}t}}\sqrt{\frac{3H_{0}}{3H_{0}C_{1}-e^{-3H_{0}t}}},
\end{equation}

\begin{equation}
\psi\left(t\right)=\frac{C_{1}}{\mu C_{2}^{4}}\left(\frac{\Lambda C_{2}^{4}}{6}+\mu C_{1}^{2}\right)e^{4H_{0}t}-\frac{2}{3\mu H_{0}C_{2}^{4}}\left(\frac{\Lambda C_{2}^{4}}{12}+\mu C_{1}^{2}\right)e^{H_{0}t}+\frac{1}{8\mu C_{2}^{2}}+\frac{C_{1}}{9H_{0}^{2}C_{2}^{4}}e^{-2H_{0}t}.
\end{equation}

\subsection{Inflation scenario}

We will consider a toy inflationary scenario defined by the following
Hubble rate \citep{Odintsov:2015SingularInflationary}:

\begin{equation}
H\left(t\right)=c_{0}+f_{0}\left(t-t_{s}\right)^{\alpha},\label{eq: Hubble_inf}
\end{equation}
with the assumption that $c_{0}\gg f_{0}$ and also for the cosmic
times near the inflationary era, it holds true that $c_{0}\gg f_{0}\left(t-t_{s}\right)^{\alpha}$
for $\alpha>0$. So in effect, near the time instance $t\simeq t_{s}$,
the cosmological evolution is nearly de Sitter. Also, the type IV
singularity occurs at $t=t_{s}$, as it can be seen from Eq.(\ref{eq: Hubble_inf}).
Particularly, the singularity structure of the cosmological evolution
(\ref{eq: Hubble_inf}) is determined from the values of the parameter
$\alpha$, and for various values of $\alpha$ it is determined as
follows:
\begin{itemize}
\item $\alpha<-1$ corresponds to the type I singularity,
\item $-1<\alpha<0$ corresponds to type III singularity,
\item $0<\alpha<1$ corresponds to type II singularity,
\item $\alpha>1$ corresponds to type IV singularity.
\end{itemize}
The detailed classification of finite time cosmological singularities
can be found in \citep{nojiri2005properties}. 

Nevertheless, obtaining exact expressions for the scalar functions
$\phi\left(t\right)$ and $\psi\left(t\right)$ within the framework
of the Hubble parameter in Eq.(\ref{eq: Hubble_inf}) presents challenges.
Therefore, exploring specific cases becomes beneficial. For instance,
to achieve viable inflation, consider the following constraints: $c_{0}=H_{0}$,
$f_{0}=-M^{2}/6$, $t_{s}=t_{i}$ and $\alpha=1$. Under these conditions,
the Hubble rate as a function of cosmic time is given by:

\begin{equation}
H\left(t\right)=H_{0}-\frac{M^{2}}{6}\left(t-t_{i}\right),\label{eq: Hubble_starobinski}
\end{equation}
where $H_{0}$, $M$, and $t_{i}$ are arbitrary constants. This framework
is known the $R^{2}$ Starobinsky inflation model \citep{Starobinsky:1980ANew,Barrow:1988Inflation,Odintsov:2015SingularInflationary}
which has non-singular characteristic. Actually $t_{i}$ represents
the beginning of inflation, which can be considered as the horizon
crossing of the large scale mode ($\sim0.05Mpc^{-1}$), and moreover,
$H_{0}$ the value of the Hubble rate at $t_{i}$. Thereby unlike
to the de-Sitter case, the Starobinsky inflation provides an exit
of inflation. In this particular model, since the study of this model
is done for early times, we can approximate this scale factor as follows,
\begin{equation}
a\left(t\right)\simeq a_{0}e^{H_{0}t+\frac{M^{2}}{6}tt_{i}}.
\end{equation}
Therefore, the scalar functions reduce to the following forms,
\begin{equation}
\phi\left(t\right)=\frac{C_{2}}{a_{0}e^{H_{0}t+\frac{M^{2}}{6}tt_{i}}}\sqrt{\frac{M^{2}t_{i}+6H_{0}}{-2e^{H_{0}t+\frac{M^{2}}{6}tt_{i}}+C_{1}a_{0}^{3}\left(M^{2}t_{i}+6H_{0}\right)}},
\end{equation}

\begin{eqnarray}
\psi\left(t\right) & = & \frac{e^{-\frac{t\left(M^{2}t_{0}+6H_{0}\right)}{3}}}{\mu C_{1}C_{2}^{4}a_{0}^{2}\left(M^{2}t_{i}+6H_{0}\right)^{2}}\Biggl\{-4C_{1}a_{0}^{3}\left(M^{2}t_{i}+6H_{0}\right)\left(\mu C_{1}^{2}-\frac{\Lambda C_{2}^{4}}{12}\right)e^{\frac{t\left(M^{2}t_{i}+6H_{0}\right)}{2}}\nonumber \\
 &  & +\frac{C_{1}C_{2}^{2}a_{0}^{2}}{8}\left(M^{2}t_{i}+6H_{0}\right)^{2}e^{\frac{t\left(M^{2}t_{i}+6H_{0}\right)}{3}}+\mu C_{1}^{4}a_{0}^{6}\left(M^{2}t_{i}+6H_{0}\right)^{2}e^{t\left(M^{2}t_{i}+6H_{0}\right)}-\frac{2\Lambda C_{2}^{4}}{3}+4\mu C_{1}^{2}\Biggr\}.
\end{eqnarray}
Furthermore, we can easily say that it is possible to find different
inflationary models by using different limitation within Eq.(\ref{eq: Hubble_inf}).

\subsection{Non-singular bouncing cosmology}

The non-singular bouncing cosmology proposes an alternative to the
traditional Big Bang model by suggesting that the universe undergoes
cycles of contraction and expansion without ever reaching a singularity.
Let us continue with a bouncing scale factor of the form \citep{Cai:2011MatterBounce}
\begin{equation}
a\left(t\right)=a_{B}\left(1+\frac{3}{2}\sigma t^{2}\right)^{\frac{1}{3}},\label{eq: scale_bounce}
\end{equation}
where $a_{B}$ represents the scale factor at the bouncing point,
and $\sigma$ is a positive parameter that describes the rate at which
the bounce occurs. This ansatz demonstrates the bouncing behavior,
corresponding to matter-dominated contraction and expansion. In this
scenario, $t$ ranges from $-\infty$ to $+\infty$, with $t=0$ being
the bounce point. By using the scale factor in (\ref{eq: scale_bounce}),
$\phi\left(t\right)$ and $\psi\left(t\right)$ read as,
\begin{equation}
\phi\left(t\right)=\frac{\sqrt{3}C_{2}}{\left(1+\frac{3}{2}\sigma t^{2}\right)^{\frac{1}{3}}}\sqrt{\frac{a_{B}\sqrt{\sigma}}{3C_{1}a_{B}^{3}\sqrt{\sigma}+\sqrt{6}\mathcal{A}}},
\end{equation}

\begin{equation}
\psi\left(t\right)=-\frac{1}{a_{B}^{2}\mu\sigma C_{2}^{4}}\left(3C_{1}a_{B}^{3}\sqrt{\sigma}+\sqrt{6}\mathcal{A}\right)^{2}\left(1+\frac{3}{2}\sigma t^{2}\right)^{\frac{4}{3}}\left(\sigma C_{2}^{2}a_{B}^{2}\int\mathcal{B}\text{d}t-\frac{C_{1}\mu}{9}\right),\label{eq: psi_bounce}
\end{equation}
where we have defined $\mathcal{A}\left(t\right)=\arctan\left(\frac{\sqrt{6\sigma}}{2}t\right)$
and the function $\mathcal{B}\left(t\right)$ inside the integral
in Eq.(\ref{eq: psi_bounce}) is given by
\begin{eqnarray}
\mathcal{B}\left(t\right) & = & \frac{27}{4\left(3C_{1}a_{B}^{3}\sqrt{\sigma}+\sqrt{6}\mathcal{A}\right)^{5}\left(1+\frac{3}{2}\sigma t^{2}\right)^{\frac{4}{3}}}\Biggl\{2a_{B}C_{2}^{2}\Lambda\biggl[\frac{4}{27}\sqrt{6}\mathcal{A}^{3}+\frac{4}{3}a_{B}^{3}C_{1}\sqrt{6\sigma}\mathcal{A}^{2}\nonumber \\
 &  & +\frac{2}{3}a_{B}^{6}C_{1}^{2}\sigma\sqrt{6}+\frac{2}{3}a_{B}^{9}C_{1}^{3}\sigma^{3/2}\biggr]\left(1+\frac{3}{2}\sigma t^{2}\right)^{\frac{4}{3}}+\frac{8\sqrt{6}\sigma\mathcal{A}^{3}t}{9}+4\sqrt{\sigma}\left(\frac{1}{3}+2C_{1}a_{B}^{3}\sigma t\right)\mathcal{A}^{2}\nonumber \\
 &  & +4\sqrt{6}a_{B}^{3}C_{1}\sigma\left(\frac{1}{3}+\sigma C_{1}a_{B}^{3}t\right)\mathcal{A}+2\sigma^{3/2}C_{1}^{2}a_{B}^{6}\left(1+4C_{1}a_{B}^{3}\sigma t\right)\Biggr\}.
\end{eqnarray}

\subsection{Cyclic cosmology}

Cyclic cosmologies propose a universe undergoing continuous cycles
of expansion and contraction. These scenarios challenge the conventional
notion of a one-time beginning and suggest that the universe could
be eternal, with each cycle resetting the conditions for the next.
This perspective aligns with some philosophical and theoretical considerations
about the nature of time and existence. To demonstrate cyclicity in
this scenario, we begin by imposing a suitable form of $a\left(t\right)$
that corresponds to cyclic behavior. Specifically, we consider a straightforward
yet broadly applicable example: a cyclic universe with an oscillatory
scale factor of the form \citep{Cai:2011Cyclic}

\begin{equation}
a\left(t\right)=a_{c}+A\sin\left(\omega t\right).\label{eq: scale_cyc}
\end{equation}
where $A$ and $\omega$ are constants, and the non-zero constant
$a_{c}$ is included to remove potential singularities from the model.
In this situation, the parameter $t$ ranges from $-\infty$ to $+\infty$,
with $t=0$ being a specific moment without any particular physical
meaning. Finally, note that the bounce occurs at $a_{B}\left(t\right)=a_{c}-A$.
For a simple solution let's set $C_{1}=0$ then the scalar functions
transform to
\begin{equation}
\phi\left(t\right)=-\frac{\left(A-a_{c}\right)^{\frac{5}{4}}\sqrt{2\omega C_{2}}}{C_{2}\mathcal{B}},
\end{equation}

\begin{eqnarray}
\psi\left(t\right) & = & -\frac{2\Lambda\mathcal{B}^{2}}{3\mu\omega\left(A-a_{c}\right)^{\frac{5}{4}}}\Biggl\{4\left(A^{2}+2a_{c}^{2}\right)\mathcal{B}^{2}\text{arctanh}\left(\frac{\left(A-a_{c}\right)^{3}\left(A+a_{c}\right)^{2}}{3\mu\omega\left(A^{2}-a_{c}^{2}\right)}\tan\left(\frac{\omega t}{2}\right)\right)\nonumber \\
 &  & +A\sqrt{A^{2}-a_{c}^{2}}\cos\left(\frac{\omega t}{2}\right)\sin\left(\frac{\omega t}{2}\right)\left(-3a_{c}\mathcal{B}+A^{2}+a_{c}^{2}\right)\Biggr\}.
\end{eqnarray}
Here, we introduce the following time-dependent function for simplicity,
\begin{equation}
\mathcal{B}\left(t\right)=2A\cos\left(\frac{\omega t}{2}\right)^{2}-A+a_{c}.\label{eq: B_cyc}
\end{equation}

\section{Conclusion\label{sec:Conclusion}}

In this paper, we have explored the gauge theory of gravity based
on the Maxwell-Weyl group and its potential for a more comprehensive
cosmological model. We carried out our investigation by applying an
analogous methodology to that employed in the Einstein-Yang-Mills
cosmology, we have established a cosmological model that incorporates
generalized Friedmann equations including the effective cosmological
constant and the contribution of additional scalar fields originating
from the Maxwell symmetry in Eqs.(\ref{eq: Friedmann_00})-(\ref{eq: Friedmann_acceleration}).

The obtained result extends the conventional cosmological solution
derived from the Weyl gauge theory of gravity, as described in Rosen
\citep{Rosen:1982Weyl}, by introducing the Maxwell symmetry under
a particular gauge constraint expressed in Eq.(\ref{eq: cosmic_gauge}).
Moreover, the present findings suggest that the Maxwell gauge field
may play a significant role in the acceleration of the universe, as
indicated in Eq.(\ref{eq: Friedmann_acceleration}), and these new
terms may be interpreted as a kind of dark energy. In this context,
because of the time dependence of the functions $\phi\left(t\right)$
and $\psi\left(t\right)$, this theory has a potential to explain
dynamical form of dark energy in time (for example the cyclic universe
model \citep{Steinhardt:2002CosmicEvolution} or quintessence dark
energy \citep{Caldwell:1998CosmologicalImprint}). In addition, the
cosmic gauge in Eq.(\ref{eq: cosmic_gauge}) provide a geometric explanation
for the Dirac scalar field. Thus, our theory arises as a scalar-tensor
theory of gravitation in which the scalar and tensor fields are of
geometrical significance similar to Lyra's gravity. Furthermore, it
is worth noting that in a specific limit, the modified Friedmann equations
presented in this study reduce to the well-known cosmological equations
of general relativity with the cosmological constant, (see Eq.(\ref{eq:Friedmann_classical})).

In Section \ref{sec:Applications-for-various}, we explored various
cosmological scenarios within the context of the generalized gravitation
model given in this study. According to the results in this section,
our model successfully incorporates power law evolution, de Sitter-like
expansion, inflationary dynamics, non-singular bounce cosmologies,
and cyclic cosmologies. The power law evolution and de Sitter-like
expansion solutions provide powerful frameworks for the accelerated
expansion of the universe, both during inflation and in the current
dark energy-dominated epoch in the context of the standard Big Bang
evolution of the universe. For the inflation era, we also considered
the Starobinsky kind of inflation which has an graceful exit, and
also, the observable indices (like the spectral index and the tensor-to-scalar
ratio) are in agreement with the Planck data \citep{Planck:2018}.
The incorporation of non-singular bounce and cyclic cosmologies offer
an alternative to the traditional hot Big Bang model by avoiding the
initial singularity. This approach suggests a pre-Big Bang phase,
potentially connected to earlier cycles or different universes, thus
offering a new perspective on the universe's origin. Each of these
solutions addresses specific epochs and phenomena in the universe's
evolution, from the early universe to potential future behaviors.

Furthermore, the cyclic models in Eqs.(\ref{eq: scale_cyc})-(\ref{eq: B_cyc})
and the conformal characteristics of the Maxwell-Weyl gravity open
the door for potential connections with Conformal Cyclic Cosmology
(CCC) \citep{penrose:2010cyclesBook,Penrose:2012CCC}, which posits
that the universe iterates through infinite aeons, each beginning
with a Big Bang and ending in an accelerated, de Sitter-like phase.
Future research could explore how our cyclic solutions might interface
with the principles of CCC, offering a deeper understanding of the
universe's temporal dynamics and the transition mechanisms between
cycles.

Consequently, we have demonstrated how the additional contributions
come from the Maxwell-Weyl gauge theory of gravity alter the standard
picture of hot Big Bang cosmology by analyzing solutions for various
cosmological phenomena. By incorporating the considered diverse cosmological
phenomena, our model offers a comprehensive framework that enhances
our understanding of the universe's evolution and addresses key questions
in modern cosmology.

\section*{Conflicts of Interest}

The authors declare no conflicts of interest.

\section*{Data Availability Statement}

No new data were created or analysed in this study.

\section*{Declaration of generative AI and AI-assisted technologies in the
writing process}

During the preparation of this work the author used GPT-3.5 in order
to improve readability of the text. After using this tool/service,
the author reviewed and edited the content as needed and take full
responsibility for the content of the publication.
\begin{acknowledgments}
This study is supported by the Scientific and Technological Research
Council of Turkey (TUB\.{I}TAK) under grant number 2219.
\end{acknowledgments}

\bibliography{MW_C}

\end{document}